\documentclass[preprint,12pt]{aastex}

\shorttitle{A Correlation between Mass Ratio and Period Ratio in
  Planetary Systems} \shortauthors{Mazeh \& Zucker}

\begin{document}

\title{A Possible Correlation between Mass Ratio and Period Ratio in
  Multiple Planetary Systems}

\author{Tsevi Mazeh\altaffilmark{1} and Shay Zucker\altaffilmark{2}}

\altaffiltext{1}{School of Physics and Astronomy, Raymond and Beverly Sackler
Faculty of Exact Sciences, Tel Aviv University, Tel Aviv, Israel}
\altaffiltext{2}{Department of Geophysics and Planetary Sciences, Raymond and 
Beverly Sackler Faculty of Exact Sciences, Tel Aviv University, Tel
Aviv, Israel}
\email{mazeh@wise.tau.ac.il; shay@wise.tau.ac.il} 

\begin{abstract}
  
  We report on a possible correlation between the mass ratio and
  period ratio of pairs of adjacent planets in extra-solar planetary
  systems. Monte-Carlo simulations show that the effect is significant
  to level of 0.7\%, as long as we exclude two pairs of planets whose
  periods are at the 1:2 resonance. Only the next few multiple systems
  can tell if the correlation is real.

\end{abstract}


\section{The correlation between the mass ratio and the period ratio}

As of March 2003, $101$ extra-solar giant planets have been
discovered, with minimum masses between $0.12$ and $15$ Jupiter masses
and orbital periods between $2.986$ and $5360$ days\footnote{see
  http://exoplanets.org/almanacframe.html}. The periods and the masses
of the planets are apparently correlated. The emerging population has
shown indications for a correlation that corresponds to a paucity of
massive planets with short orbital periods. Furthermore, this
correlation does not appear in the population of planets that have
been found in stellar binary systems (Zucker \& Mazeh 2002).

The known extra-solar planets include a special subgroup of $22$
planets found in $10$ multiple systems (Fischer et al.\ 2003), two of
which consist of $3$ planets ($\upsilon$\,And and 55\,Cnc). In this
{\it Letter} we focus on the masses and orbital periods of the planets
found in the multiple systems, and present a distinctive correlation
that characterizes this subsample.

The multiple systems provide us with a unique feature --- the mass and
period {\it ratios} between planets in adjacent orbits, ratios that
can reflect general characteristics of the multiple systems. Although
only the minimum masses are known for each planet, we assume the
orbital inclinations of the planets in the same multiple system are
similar, and therefore the ratio of the minumum masses is very close
to the ratio of the actual masses. We therefore studied here the
correlation between the (minimum) mass and the period ratios of all
pairs of planets in adjacent orbits.

For each multiple system with two planets we derived one mass ratio
and one orbital-period ratio. For each of the two systems with three
known planets, we derived two sets of ratios --- one set of ratios
between the intermediate planet and the innermost one, and one set of
ratios between the outermost and the intermediate one.  Altogether,
we have $12$ such pairs of extra-solar planets. In Figure
\ref{fig1} we plotted their mass ratios as a function of their period
ratios.

The Solar System includes two giant planets, Jupiter and Saturn,
within the mass range of the known extra solar planets.  We added
an open circle in the Figure to represent the mass and period ratios
of the Saturn/Jupiter pair. The mass and period ratios are very
similar to those of 47 UMa, as already pointed out by Fischer et al.\ 
(2002). 

The Figure shows an intriguing correlation between the two ratios.
Except for two points that lie exactly at the 1:2 orbital-period
resonance, all points seem to fit a straight line in a log-log plot.
Considering the extra-solar planets alone, the correlation between the
logarithms of the two ratios is $0.9415$ and the best-fit line, which
is also plotted in the Figure, has a slope of $0.92 \pm 0.10$ ---
suspiciously close to unity. When the fit includes the solar-system
point, the correlation rises to $0.9498$.

\section{Significance}

The number of points in Figure \ref{fig1} is extremely small. We have
altogether only $12$ points (excluding Saturn/Jupiter), out of which,
we claim, $2$ points should be excluded, because of their unique
period ratio.  On top of that, the data are subject to a strong
selection effect that thwarts the detection of extra-solar planets
with small masses and long periods. On the other hand, the correlation
is intriguingly high, even for such a small number of points.

To estimate the significance of our findings we performed two
randomization tests. In the first one we have used the masses and
orbital periods of all known planets, a set that is supposedly subject
to similar observational selection effect. We chose at random $8$
pairs and $2$ triples of orbits from the $101$ extra-solar orbits, and
calculated the correlation between their mass ratios and period
ratios. We removed the two pairs that maximized the correlation for
the remaining $10$ pairs. Out of $1,000,000$ random choices, only
$7921$ (0.8\%) yielded correlations higher than that obtained for the
true data of the extra-solar pairs of planets.

The Monte-Carlo simulation described above shows the significance of
the linear relation for the extra-solar planets alone. As we have seen
in the previous section, the Saturn/Jupiter pair seems to agree with
the same rule. In order to take account of this fact, we repeated the
simulations, this time adding this pair to the randomly drawn $10$
pairs and $2$ triples, before rejecting two points.  The significance
implied by these modified simulations is $0.3\%$!

Note, however, that the selection effects of the single planets could
have been quite different from the selection effects of the multiple
systems. This might be the case mainly because once a planet has been
discovered in a system, the observation strategy changes, leading to a
different set of biases when additional planets are discovered in the
same system.  In general, large spread of a few radial velocity
measurements of the stars in the sample is the first hint of the first
planet, which usually entails further extensive series of
observations. Large residuals relative to the derived orbit are a sign
for a second planet, which can be the drive for additional
observations that might detect and measure the orbital motion of the
second (and third) planet.

Furthermore, the difficulty of extracting multiple signatures from the
same data set probably introduces certain biases when looking for low
mass planets in multiple systems. It might be more difficult to
identify a low mass planet in a system which already contains a large
planet if the orbital periods are similar rather than if they differ
by a factor of 10 or more.

To try and overcome these difficulties we ran another randomization
test in which we considered only the planets found in multiple
systems.  Again, we chose at random $8$ pairs and $2$ triples of
orbits only from the $22$ extra-solar orbits found at the multiple
systems, and calculated the correlation between their mass ratios and
period ratios. The period and mass distributions of the inner and
outer planet in our simulated population are very similar to those of
the actual pairs. For each set of simulated pairs we removed the two
pairs that maximized the correlation for the remaining $10$ pairs. Out
of $1,000,000$ random choices, only $6543$ (0.65\%) yielded
correlations higher than that obtained for the true data of the
extra-solar pairs of planets.  In order to visualize the distribution
of the simulated correlation coefficients, we applied the Fisher's
z-transformation (e.g., Kendall \& Stuart 1973) to the simulated
coefficients, resulting in an approximately symmetrical distribution.
The result is depicted in Figure \ref{fig2}.  The correlation obtained
for the true data is marked by an arrow.

Note that our simulations rejected the points whose removal optimally
improved the correlation.  We therefore expect the resulting
correlation to be relatively high. In the true data the rejected
points shared the property of lying exactly on the 1:2 resonance.  We
therefore feel our claim for a relatively high significance is well
established. 

{\it Obviously, the reality of the correlation can be proved only by
 additional multiple systems,} hopefully to be found in the near future.

\section{Discussion}

\begin{figure}
\plotone{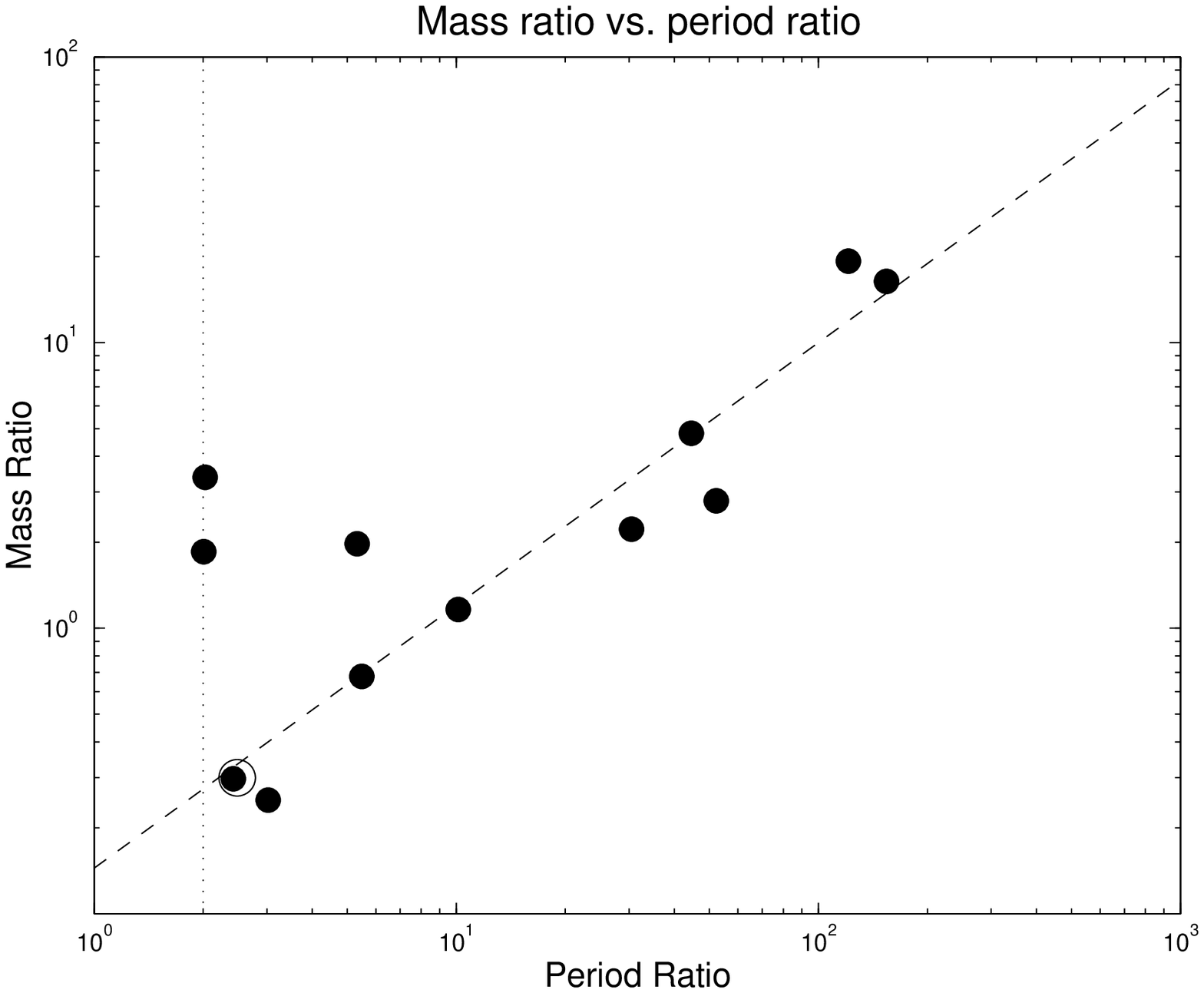}
\caption
{The mass and period ratios of the 12 extra-solar adjacent planet
pairs. The open circle represents the Saturn/Jupiter pair.
The dashed line represents the best-fit linear relation for the
extra-solar planets, not including Saturn/Jupiter point. 
The vertical dotted line represents the 1:2 resonance.
\label{fig1}}
\end{figure}

\begin{figure}
\plotone{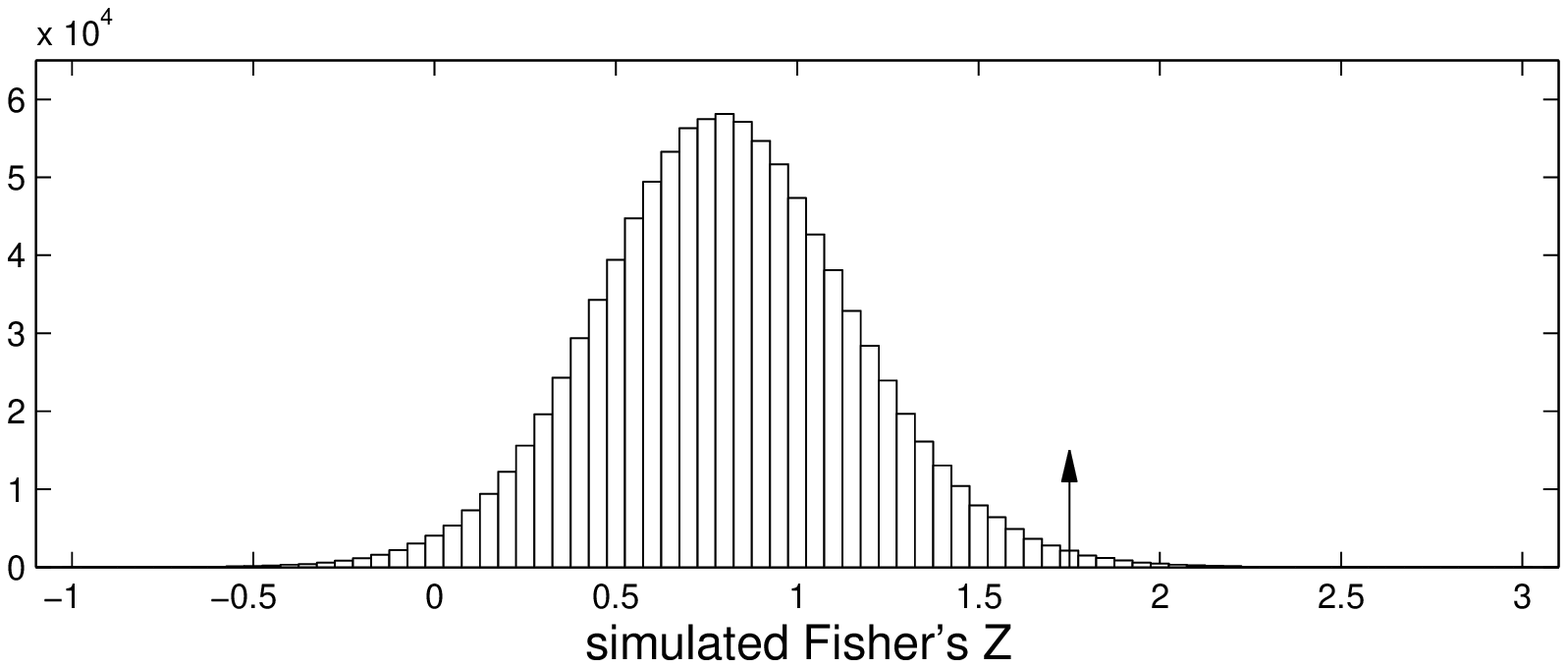}
\caption
{The distribution of $10^6$\ simulated correlation coefficients (see
text for details of the simulation).  The correlation coefficients
were transformed using the Fisher's z-transformation.  The arrow
corresponds to the correlation coefficient of the true data.
\label{fig2}}
\end{figure}

Figure \ref{fig1} shows that multiple systems with a large period
ratio (and therefore large orbital-radius ratio) show also a large
ratio between their masses, with the more massive planet on the
outside.  Within the migration paradigm (e.g., Goldreich \& Tremaine
1980; Lin, Bodenheimer \& Richardson 1996), the present orbital radii
of the planets are substantially smaller than the distance of their
formation sites from their parent stars. Therefore the correlation we
found, if verified, could be the result of some correlation between
the migration range of a planet and it's mass. Massive planets might
migrate slower (e.g., Ward 1997; Trilling et al.\ 1998; Nelson et al.\ 
2000) and therefore are left far away when the disc evaporates. A
similar effect could have caused the paucity of the massive planets
with short periods (Zucker \& Mazeh 2002).  If this is true, any point
that represents a pair of planets in our parameter space slides to the
right during migration, when the period of the smaller planet gets
shorter.

Planets that find themselves in a 1:2 orbital resonance interact
strongly with each other. The interaction could keep the period ratio
stable against different migration rates that tend to push the smaller
planet further in (e.g., Snellgrove, Papaloizou \& Nelson 2001).
Therefore we suggest that if the migration scenario is responsible for
the correlation we find, the pairs of planets with the 1:2 resonance
did not succumb to this process, and therefore stayed at this
resonance despite their large mass ratios.

The above general considerations do not explain why the correlation we
found holds for more than two decades, and in particular why the
exponent in the power law is so close to unity. Any theory of
planetary formation and/or migration would have to explain these
findings, if verified.

One other possibility to interpret the possible correlation is to
assume the present periods reflect the original distances of the
formation sites of the planets from their parent stars.  Thus, the
correlation we found may be related to a correlation between the
location of the formation site of a planet and its mass.  After all,
the larger the disc radius at the formation site, the more mass is
available for planetary accretion within a certain fraction of the
planet original radius. 

Interestingly enough, a similar idea has been proposed already by
Laskar (2000), at a paper titled ``On the Spacing of Planetary
Systems''. In that paper, he suggested a power-law relation $$
{{m_1}\over{m_2}}=\Bigl({{a_1}\over{a_2}}\Bigr)^{(2p+3)/6}\ , $$ which
translates into a power-law relation between the mass and the period
ratios. However, Laskar theory differs from the present data in two
crucial details. Laskar considered cases where $p$ is between $0$ and
$-3/2$, which translates into an exponent between $1/3$ and $0$ for
the mass ratio --- period ratio relation. The present data suggest an
exponent which is close to unity. Furthermore, extension of Laskar
relation goes, naturally, through the $(m_2/m_1,p_2/p_1) = (1,1)$
point, whereas ours does not. Therefore, Laskar theory by itself
cannot explain the possible relation we suggest here, and we probably
need a combination of models for formation and migration.

The present paper suffers from two drawbacks. First, although we have
shown the effect seems to be statistically significant, the number of
points in Figure \ref{fig1} is still quite small.  A few additional
planetary systems are critically needed in order to establish the reality
of the effect.  Second, the authors do not claim to fully understand
the possible correlation. If the effect is established in the future,
we will need theoretical studies to understand the mechanism behind
it.

\begin{acknowledgements}
  
  We wish to thank M.\ Holman and the referee, J.\ Chambers, for
  important comments and advice. This work was supported by the
  Israeli Science Foundation (grant no.\ 40/00). S.Z.\ is grateful for
  partial support from the Jacob and Riva Damm Foundation.

\end{acknowledgements}

\begin{references}

\reference{} Fischer, D.~A., Marcy, G.~W., Butler, R.~P., Laughlin, G., Vogt, \& S.~S.\ 
  2002, \apj, 564, 1028

\reference{} Fischer, D.~A., et al.\ 2003, \apj, 586, 1394

\reference{} Goldreich, P., \& Tremaine, S.\ 1980, \apj, 241, 425

\reference{} Kendall, M.~C., \& Stuart, A.\ 1973, The Advanced Theory of
  Statistics, 3rd Edition (London: Charles Griffin \& Co.)

\reference{} Laskar, J.\ 2000, \prl, 84, 3240

\reference{} Lin, D.~N.~C., Bodenheimer, P., \& Richardson, D.~C.\ 1996, \nat, 380, 606

\reference{} Nelson, R.~P., Papaloizou, J.~C.~B., Masset, F., \& Kley,
W.\ 2000, \mnras, 318, 18

\reference{} Snellgrove, M.~D., Papaloizou, J.~C.~B., \& Nelson,
R.~P. 2001, \aap, 374, 1092

\reference{} Trilling, D.~E., Benz, W., Guillot, T., Lunine, J.~I.,
Hubbard, W.~B., \& Burrows, A.\ 1998, \apj, 500, 428

\reference{} Ward, W.~R.\ 1997, Icarus, 126, 261

\reference{} Zucker, S., \& Mazeh, T.\ 2002, \apjl, 568, L113


\end{references}
\end{document}